\documentclass[12pt]{article}
\usepackage{amsmath,amssymb,amscd,cite}
\usepackage{soul,color}
\newtheorem{thm}{Theorem}[section]
\newtheorem{prop}[thm]{Proposition}
\newtheorem{lem}[thm]{Lemma}
\newtheorem{cor}[thm]{Corollary}

\newtheorem{defi}[thm]{Definition}
\newcommand{\pf}{{\bf Proof. \ }}
\newcommand{\qed}{\hfill $\blacksquare$ \\}

\font\msbm=msbm10 at 12pt
\newcommand{\Z}{\mbox{\msbm Z}}

\newcommand{\F}{\mbox{\msbm F}}
\newtheorem{rem}[thm]{Remark}
\newtheorem{ex}[thm]{Example}

\newcommand{\ord}{\mbox{ord}}
\date{}

\begin{document}

\title{On Isodual Cyclic Codes over Finite Chain Rings}
\author{Aicha Batoul, Kenza Guenda, T. Aaron Gulliver and Nuh Aydin\thanks{A. Batoul and K. Guenda
are with the Faculty of Mathematics, USTHB, University of Science and
Technology of Algiers, Algeria.
T. Aaron Gulliver is with the Department of Electrical and Computer Engineering,
University of Victoria, PO Box 1700, STN CSC, Victoria, BC, Canada V8W 2Y2.
Nuh Aydin is with Department of Mathematics and Statistics,
Kenyon College, Gambier, OH 43022.
email:a.batoul@hotmail.fr, kenguenda@gmail.com,
agullive@ece.uvic.ca, aydinn@kenyon.edu.}}
 \maketitle

\begin{abstract}
In this paper, cyclic isodual codes over finite chain rings are investigated.
These codes are monomially equivalent to their duals.
Existence results for cyclic isodual codes are given based on the generator polynomials,
the field characteristic, and the length.
Several constructions of isodual and self-dual codes are also presented.\\

\end{abstract}
\section{Introduction}
A code which is equivalent to its dual is called an isodual code.
Here we only consider monomial equivalence, which is the most important.
For some parameters, one can prove that there are no cyclic self-dual codes over finite chain
rings~\cite{aicha2011,BGGP}, whereas isodual codes can exist.
Isodual codes are important because they are related to lattices.
Recently, isodual cyclic codes over finite fields were constructed from duadic codes~\cite{BGGI}.
The purpose of this paper is to extend the concept of duadic codes to finite chain rings and to
extend the construction of isodual codes in~\cite{BGGI} to finite chain rings.
Note that duadic codes over $\Z_4$ were presented by Langevin et al. \cite{Langevin},
over $\F_2+u\F_2$ by Ling et al. \cite{SanLing}, and over $\Z_{2k}$ by Bachoc et al. \cite{bachoc}.

The remainder of this paper is organized as follows.
Section 2 provides some preliminary results.
In Section 3, the structure of cyclic codes of length $2^am$ over finite chain rings is presented.
Section 4 gives conditions on the existence of isodual cyclic codes over finite chain rings,
and several constructions are presented.
Some constructions of isodual cyclic codes over finite chain rings are provided in Section 5 using the lifts of duadic codes over the residue field.

\section{Preliminaries}

In this section, we summarize some necessary results from~\cite{permounth,G-G,Ana}.
A finite chain ring $R$ is a finite commutative ring with identity $1 \neq 0$ and maximal principal ideal generated by a nilpotent element $\gamma \in R$.
 The residue field of $R$ is $\frac{R}{\langle\gamma\rangle}$ which is denoted by $K$.

The natural surjective ring morphism $(-)$ is given by
\begin{equation}
\label{eq:over}
\begin{split}
-:  R&  \longrightarrow K\\
a &\longmapsto \overline{a}=a \bmod \gamma.
\end{split}
\end{equation}

A code $C$ of length $n$ is called a linear code over a finite chain ring $R$ if it is a submodule of $R^n$.
Here, all codes are assumed to be linear.
A code $C$ is said to be cyclic if
\[
(c_{n-1},c_{0},\ldots,c_{n-2}) \in C \text{ whenever }(c_{0},c_{1}, \ldots, c_{n-1}) \in C.
\]
We attach the standard inner product to $R^n$
\[
[{v},{w}] = \sum v_iw_i,
 \]
for $v=(v_0,v_1, \ldots, v_{n-1})$ and $w=(w_0,w_1, \ldots, w_{n-1}) \in R^n$.
The dual code $C^\perp$ of $C$ is defined as
\begin{equation}
C^\perp=\{ {v} \in R^n \ | \ [{v},{w}]= 0 {\rm \ for\
all\ } {w} \in C\}.
\end{equation}
If $C \subseteq C^\perp$, the code is said to be self-orthogonal, and if $C=C^\perp$, the code is self-dual.
%

In this paper, the notation $q=\square \bmod n$
means that $q$ is a quadratic residue modulo $n$.
For a prime power $q$ and integer $n$ such that $\gcd(q,n)=1$, we denote by $\ord_n(q)$
the multiplicative order of $q$ modulo $n$.
This is the smallest integer $l$ such that $q^l\equiv 1 \bmod n$.

Suppose that $f(x)= a_0+a_1x+\ldots +a_rx^r$ is a polynomial of degree $r$
with $f(0)= a_0$ a unit in $R$.
Then the monic reciprocal polynomial of $f(x)$ is
\[
f^*(x)= f(0)^{-1}x^rf(x^{-1})= a_0^{-1}(a_r+a_{r-1}x+\ldots +a_0x^r).
\]
If a polynomial is equal to its reciprocal, then it is called a self-reciprocal polynomial.

The following lemma is easily deduced.
\begin{lem}
\label{lem:reci}
Let $f(x)$ and $g(x)$ be two polynomials in $R[x]$ with $\deg f(x) \ge \deg g(x)$.
Then the following holds.
\begin{itemize}
\item[(i)] $[f(x)g(x)]^*=f(x)^*g(x)^*$.
\item[(ii)] $[f(x)+g(x)]^*=f(x)^*+x^{\deg f- \deg g}g(x)^*$.
\item[(iii)] If $f(x)$ is monic, then $\overline{f(x)^*}=\overline{f(x)}^*$.
\end{itemize}
\end{lem}
Now we recall some definitions concerning cyclic codes over a finite field.
Let $q$ be a prime power and let $m$ be a positive odd integer such that $(m,q) =1$.
Then for $ 0\leq i< m$, the $q$-cyclotomic coset of $i \bmod m$ is defined as
\[
Cl(i)=\{iq^{l}\bmod m | l\in \mathbb{N}\}.
\]
Let $\alpha$ be a primitive $m$-th root of unity in an extension field of $\F_{q}$,
and $C$ be a cyclic code over $\F_q$ of length $m$ generated by a polynomial $f(x)$.
$C$ is uniquely determined by its defining set
$T=\{0 \leq i < m \,|\, f(\alpha^i)=0\}$.
The defining set of a cyclic code over $\F_q$ is the union of the $q$-cyclotomic cosets related to its generator polynomial.

The following theorem gives the structure of a cyclic code (not necessarily free) and its dual over a finite chain ring.
\begin{thm}(\cite{permounth})
\label{th:prince}
Let $R$ be a finite chain ring with maximal ideal $\gamma$ and index of nilpotency $e$.
Further, let $C$ be a cyclic code over $R[x]$ of length $n$ such that $(n,p)=1$, where $p$ is the characteristic of $\overline{R}$.
Then there exists a unique family of pairwise coprime polynomials $ F_i,0\leq i\leq e$ in $R[x]$ such
that $F_0 \ldots F_e=x^n-1 $,
\[C=\langle\hat{ F}_1,\gamma
\hat{F}_2, \ldots,\gamma^{e-1} \hat{F}_e \rangle \text{ and }C^{\bot}=\langle \hat{F}_{0}^{*},\gamma\hat{F}_{e}^{*},
\ldots, \gamma^{e-1}\hat{F}_{2}^{*}\rangle,
\]
where $\hat{F}_j=\frac{x^n-1}{F_j}$ for $0<j\le e$.
Moreover, we have that the ring $R[x]/( x^n-1 )$ is a principal ideal ring.
\end{thm}

\subsection{Isometries and Monomial Maps}
Let $R^{*}\,=\, R \,\setminus\, \langle\gamma\rangle$.
A monomial transformation over $R^n$ is an $R$-linear homomorphism $\tau$ such
that there exist units $\lambda_1, \ldots ,\lambda_n$ in $R^{*}$, and
a permutation $\sigma \in S_n$ such that for all
$(x_1,x_2,\ldots,x_n)\in R^n$, we have $\tau(x_1,\ldots,x_n)=
(\lambda_1 x_{\sigma (1)}, \lambda_2 x_{\sigma (2)}, \ldots, \lambda_n x_{\sigma (n)})$.
Two linear codes $C$ and $C'$ of length
$n$ are called monomially equivalent if there exists a monomial
transformation over $R^n$ such that $\tau(C)=C'$.
A weight on a code $C$ over a finite chain ring is called homogeneous
if it satisfies the following conditions:
\begin{itemize}
\item[(i)] $\forall x\in C$, $\forall u\in R^* : w(x)=w(ux)$, and
\item[(ii)] there exists a constant $\xi= \xi(w) \in \mathbb{R}$ such that
\[
\sum_{x\in U} w(x) =\xi|U|,
\]
where $U$ is any subcode of $C$.
\end{itemize}
A linear morphism $f: R \longmapsto R$ is called a homogeneous isometry if it
is a linear homomorphism which preserves the homogeneous weight.
\begin{lem}
\label{lem:greferath}(\cite{greferath})
Let $R$ be a finite chain ring, $C$ a linear code over $R$, and $\phi: C \longmapsto R^n$ an embedding.
Then the following are equivalent:
\begin{itemize}
\item[(i)] $\phi$ is a homogeneous isometry, and
\item[(ii)] $C$ and $\phi(C)$ are monomially equivalent.
\end{itemize}
\end{lem}
Here whenever two codes are said to be equivalent it is meant that they are monomially equivalent.

The function $\mu_a$ defined on $\Z_n = \{ 0,1,\ldots,n-1\}$ by
$\mu_a(i) \equiv ia \bmod n$ is a permutation of the coordinate
positions $\{0, 1,2,\ldots,n-1\}$ and is called a multiplier.
Multipliers can also act on polynomials in $R[x]$ and this gives the following ring automorphism
\begin{equation}
\begin{array}{ccl}
\label{eq:ling}
\mu_a:R[x]/(x^n-1) &\longrightarrow &
 R[x]/(x^n-1)\\
 f(x)&\mapsto & \mu_a(f(x)) =
f(x^a).
\end{array}
\end{equation}

\subsection{Galois Extensions of Finite Chain Rings}
A cyclic code over a finite field can be defined by its defining set, i.e. the set of roots of its generator polynomial.
In this section, this definition is extended to cyclic codes over finite chain rings.
This shows that in some cases we can know the Hensel lift of a polynomial.
In general this is not always possible, since Hensel's lemma only gives the existence of a lift polynomial.
Let $R$ be a finite chain ring with residue field $\F_q$ where $\F_{q^s}$ is the splitting field of $x^n-1$ over $\F_q$
with $s=\ord_n(q)$.
Further, let $f(x)\in \F_q[x]$ be a primitive polynomial of degree $s$.
Then since $(q^s-1,q)=1$, there exists a unique basic irreducible polynomial $g(x)\in R[x]$ such that $\overline{g}(x)=f(x)$.

Consider the Galois extension of $R$ denoted by $S\simeq \frac{R[x]}{(g(x))}$.
Since $g$ is irreducible and square free,
$S$ is separable and local.
Then from \cite[Theorem 4.2]{Mac2}, $S$ has a primitive element $\xi$ which is a root of
$g(x)$ such that $\overline{\xi}= \alpha$ is a root of $f(x)$ in $\F_{q^s}$.
The map
\begin{equation}
\begin{array}{ccl}
\label{eq:sigma}\sigma:S &\longrightarrow &S\\
 \xi &\mapsto & \sigma(\xi)=\xi^q,
\end{array}
\end{equation}
is a generator of $G_R(S)$,
the Galois group of $S$ over $R$, which is isomorphic to $G_{\F_q}(\F_{q^s})$, the Galois group of $\F_{q^s}$ over $\F_q$.
As $G_{\F_q}(\F_{q^s})$ is a cyclic group, the elements of $R$ are fixed by $\sigma$ and all its powers.
Further, since $\beta=\xi^\frac{q^s-1}{n}\in S$, the Galois extension $S$ of
$R$ contains a primitive $n$-th root of unity.
In addition, $\overline{\beta}=\alpha$ is a primitive $n$-th root of unity in $\F_{q^s}$
\begin{lem}
\label{lemma:reciproque}
With the above assumptions, let $\displaystyle{p(x)=\Pi_{i\in T}(x-\alpha^{i})}$ be a monic divisor of $x^n-1$ in
$\F_q[x]$, where $T$ is the defining set of the cyclic code $\langle p(x)\rangle$.
Then there is a unique monic factor $q(x)$ of $x^n-1$ in $R[x]$
such that $\displaystyle{q(x)=\Pi_{i\in T}(x-\beta^i)}$,
and $\overline{q}(x)= p(x)$.
\end{lem}
\pf Let $q(x)$ be the unique monic Hensel lift of $p(x)$ which is a divisor of $x^n-1$ in $\F_q[x]$,
and define
\[
\tilde{q}(x) = \Pi_{i\in T}(x-\beta^i),\,i \in \Z_n.
\]
From (\ref{eq:sigma}) we have that $\sigma(\tilde{q}(x))=\tilde{q}(x)$,
so $\tilde{q}(x)$ has coefficients from $R$.
Further $\overline{\tilde{q}(x)}= \overline{\Pi_{i\in T}(x-\beta^i)}=\Pi_{i\in T}(x-\alpha^{i})= p(x)=\overline{q}(x)$
and since $q(x)$ is unique, we have that
\[
q(x)=\Pi_{i\in T}(x-\beta^i).
\]
\qed

\section{Cyclic Codes of Length $2^am$ over $R$}
Let $R$ be a finite chain ring with residue field $\F_q$ such that $q$ is an odd
prime power, and $m$ be an odd integer such that $(m,q)=1$.
In the following we give the structure of cyclic codes of length $2^am$
where $a\geq1$ is an integer and $R^*$ contain a primitive $2^a$-th root of the unity.

We begin with the following lemma.
\begin{lem}
\label{lem:primitive}
Let $R$ be a finite chain ring with residue field $\F_{q}$ where $q$ is an odd prime power
$q=p^r$, $a\geq 1$ an integer.
Then there exists a primitive $2^a$-th root of unity $\alpha$ in $R^*$ if and only if
$q \equiv 1 \bmod 2^a$.
Further, $x^{2^a}-1=\prod_{k=1}^{2^a}(x-\alpha^k)$ in $R[x]$.
\end{lem}
\pf Since $q$ is an odd prime power, by \cite[Proposition 4.2]{aicha2012}, there exists a primitive
$2^a$-th root of unity in $R^*$ if and only if there exists a
primitive $2^a$-th root of unity in $\F_q$.
If there exists a primitive $2^a$-th root of unity $\alpha$ in $\F_q^*$,
then $\alpha^{2^a}=1$, so that $2^a$ divides $q-1$.
Conversely, if $2^a$ divides $q-1$ then there exists an integer $k$ such that $q = k2^a+1$.
If $\alpha$ is a primitive element of $\F_q^*$, then $1= \alpha^{q-1}= (\alpha^{k})^{2^a}$
and $\ord(\alpha^k)= \frac{{\rm ord}(\alpha)}{(k, {\rm ord}(\alpha))}=\frac{q-1}{(k,q-1)}=\frac{k2^a}{(k,k2^a)}=2^a$.

Let $\alpha$ be a primitive $2^a$-th root of the unity in $R^*$.
Since $(2^a,q)=1$, it must be that $\overline{\alpha}$ is a primitive $2^a$-th root of unity in $\F_q^*$
so that $x^{2^a}-1=\prod_{k=1}^{2^a}(x-\overline{\alpha}^k)$ in $\F_q[x]$.
By Lemma~\ref{lemma:reciproque} there is a one-to-one correspondence between the set of basic
irreducible polynomial divisors of $x^{2^a}-1$ in $R[x]$ and the
set of irreducible divisors of $\overline{x^{2^a}-1}$ in $\F_q[x]$.
If $x^{2^a}-1=\prod_{k=1}^{2^a}(x-a_{k})$,
then $\overline{(x-a_{k})}=(x-\overline{a_{k}})=(x-(\overline{\alpha})^{k})$.
Since $\overline{(x-\alpha^{k})}=(x-\overline{\alpha^{k}})=(x-(\overline{\alpha})^{k})$,
from the unique decomposition of $x^{2^a}-1$ in $R[x]$,
the result follows.
\qed

\begin{lem}
\label{lem:primitive2}
\hfill
\begin{enumerate}
  \item [(i)] If there exists a primitive $2^a$-th root of unity $\alpha$ in $R^*$,
  then $\alpha^{2^i}$ is a primitive $2^{a-i}$-th root of unity in $R^{*}$ for all $i\leq a$.
  \item [(ii)] Let $\alpha$ be a primitive $2^a$-th root of unity in $R^*$.
   Then $\alpha^m$ is also a primitive $2^a$-th root of unity in $R^*$.
\item [(iii)] If $a\geq 2$, then $\prod_{k=1}^{2^a}\alpha^k=1$.
\end{enumerate}
\end{lem}
\pf
By \cite[Proposition 4.2]{aicha2012}, there exists a primitive
$2^a$-th root of the unity in $R^*$ if and only if there exists a
primitive $2^a$-th root of unity in $\F_q$.
Now using \cite[Lemma 3.2]{BGGI} we have the following.\\
For part (i), for $i$, $i\leq a$, in the cyclic group $\F_q^*$,
we have $\ord(\overline{\alpha}^{2^i}) = \frac{\ord(\overline{\alpha})}{(2^i,\ord(\overline{\alpha}))}=\frac{2^a}{(2^i,2^a)}=\frac{2^a}{2^i}=2^{a-i}$.\\
For part (ii), since $(2^a,m)=1$, then
$\ord(\overline{\alpha^m})$= $\ord((\overline{\alpha})^m)= \frac{\ord(\overline{\alpha})}{(m,\ord(\overline{\alpha}))}=\frac{2^a}{(m,2^a)}=2^a$.\\
For part (iii), since $\displaystyle{(x^{2^a}-1)=\prod_{k=1}^{2^a}(x-\alpha^k)}$, then $\displaystyle{\prod_{k=1}^{2^a}\alpha^k=(-1)^{2^{a-1}}}$.
\qed
The next lemma can easily be obtained from Lemma \ref{lemma:reciproque}.
\begin{lem}
\label{lem:factorization}
Let $R$ be a finite chain ring with residue field $\F_q=\F_{p^t}$ and $m$ be an integer such that $(p,m)=1$.
Then there exist unique monic basic irreducible pairwise coprime factors $g_i(x)$, $i\in\{1,2,\ldots,r\}$, of $x^m-1$ in $R[x]$ such that
\begin{equation}
\label{eq:factorization}
x^{m}-1 =(x-1)\prod_{i=1}^rg_{i}(x).
\end{equation}
\end{lem}

\subsection{Free Cyclic Codes of Length $2^am$ over $R$}
Before giving the structure of free cyclic codes of length $2^am$ over $R$, we give the following proposition.
\begin{prop}
\label{prop:product}
Let $R$ be a finite chain ring with residue field $\F_{q}$,
$q=p^t$ be an odd prime power, and $n=2^am$ a positive
integer such that $m$ is an odd integer, $a \geq 1$ and $(m,p)=1$.
If $R^{*}$ contains a primitive
$2^a$-root of unity and $x-1$, $g_i(x)$, $1 \le i \le r$, are
the monic basic irreducible pairwise coprime factors of $x^{m}-1 $ in $R[x]$,
then
\[
x^{2^am}-1=(x^{2^a}-1) \prod_{i=1}^{r}g_{i}(\alpha^{-k}x).
\]
\end{prop}
\pf Assume that $x^{m}-1 =(x-1)\prod_{i=1}^{r}g_{i}(x)$ (so that $g_0(x)=(x-1)$).
Since $(m,p)=1$, by \cite[Theorem 4.3]{G-G} and Lemma \ref{lem:factorization} this
is the unique factorization of $x^{m}-1$ into monic basic irreducible pairwise coprime polynomials over $R$.
Let $\alpha \in R^*$ be a primitive $2^a$-th root of unity and let $1\leq k\leq2^a$.
Then
\[
\begin{array}{ccl}
(\alpha^{-k}x)^m-1&=& (\alpha^{-k}x-1)\prod_{i=1}^{r}g_{i}(\alpha ^{-k}x)\\
(\alpha^{-k})^m(x^m-(\alpha^{k})^m)& =&\alpha^{-k}(x-\alpha^k)\prod_{i=1}^{r}g_{i}(\alpha ^{-k}x)\\
(x^m-\alpha^{km})&=& \alpha^{k(m-1)}(x-\alpha^k)\prod_{i=1}^{r}g_{i}(\alpha ^{-k}x)\\
(x^m-(\alpha^{m})^k)&=& \alpha^{k(m-1)}(x-\alpha^k)\prod_{i=1}^{r}g_{i}(\alpha ^{-k}x),
\end{array}
\]
and by Lemma \ref{lem:primitive} $\alpha^{m}$ is also
a primitive $2^a$-th root of unity in $R^*$.
We have that
\[
\begin{array}{ccl}
\prod_{k=1}^{2^a}(x^m-(\alpha^{m})^k)&=&\prod_{k=1}^{2^a}\alpha^{k(m-1)}(x-\alpha^k)\prod_{i=1}^{r}g_{i}(\alpha ^{-k}x)\\
&=&\prod_{k=1}^{2^a}\alpha^{k(m-1)}\prod_{k=1}^{2^a}(x-\alpha^k)\prod_{k=1}^{2^a}\prod_{i=1}^{r}g_{i}(\alpha ^{-k}x)\\
&=&\prod_{k=1}^{2^a}\frac{\alpha^{km}}{\alpha^k}\prod_{k=1}^{2^a}(x-\alpha^k)
\prod_{k=1}^{2^a}\prod_{i=1}^{r}g_{i}(\alpha ^{-k}x)\\
&=&(x^{2^a}-1)\prod_{k=1}^{2^a}\prod_{i=1}^{r}g_{i}(\alpha ^{-k}x).
\end{array}
\]
Since $(x^{2^am}-1)=((x^m)^{2^a}-(\alpha^m)^{2^a})=\prod_{k=1}^{2^a}(x^m-\alpha^{km})$,
the result follows.
\qed

We now give the structure of free cyclic codes of length $2^am$ over $R$.

\begin{cor}
\label{cor:structure cyclic}
If $R^{*}$ contains a primitive $2^a$-root of unity $\alpha$ and $(x-1),g_i(x)$, $1\le i \le r$, are
the monic basic irreducible factors of $x^{m}-1 $ in $R[x]$, then a free cyclic code $C$ of length $n=2^am$ is generated by $\prod_{k=1}^{2^a}((x-\alpha^k)^{l_k}\prod_{i=1}^{r}g_{i}^{j_i}(\alpha ^{-k}x))$
with $1\leq l_k,j_i\leq p^s$.
\end{cor}
\pf
By \cite[Theorem 4.16]{G-G}, any free cyclic code of length $2^am$ is generated by a divisor of $x^{2^am}-1$, and
by Proposition \ref{prop:product} we have that
\[
(x^{2^am}-1)=\prod_{k=1}^{2^a} ((x-\alpha^k)\prod_{i=1}^{r}g_{i}(\alpha ^{-k}x)),
\]
and the result follows.
\qed

Next, the structure of cyclic codes (not necessarily free) of length $2^am$ over $R$ are examined.
\begin{thm}
Let $R$ be a finite chain ring with residue field $\F_{q}=\F_{p^t}$
such that $q \equiv 1 \bmod 2^a$, with $a \geq 1$ and $m$ an odd integer.
Further, let $C$ be a code of length $2^am$ over $R$.
Then $C$ is a cyclic code of length $2^am$ over $R$ if and only if
$C\simeq \bigoplus_{1\le i\le 2^a }C_i$, where $C_i$ is a cyclic code of length $m$ over $R$.
\end{thm}
\pf
Since $q\equiv 1 \bmod 2^a$, by Lemma~\ref{lem:primitive} there exists a primitive $2^a$-th root of unity
$\alpha \in R^*$ such that $\alpha^{2^a}=1$.
By Lemma \ref{lem:primitive}, $\alpha^m$ is also a primitive $2^a$-th root of unity in $ R^*$ (note that $m$ is odd),
and thus
\[
(x^{2^am}-1) = \prod_{i=1}^{2^a}(x^m-(\alpha^{m})^{i}).
\]
Since $(2^am,p)=1$, there are no repeated roots so the polynomials $(x^m-\alpha^i)$,$i\in\{1,\ldots,2^a\}$ are coprime.
Then by the Chinese Remainder Theorem we have the following ring isomorphism
\[
\frac{R[x]}{(x^{2^am}-1)} \simeq \prod_{i=1}^{2^a}\frac{R[x]}{(x^m-(\alpha^{i})^m)}.
\]
From \cite[Theorem 4.3]{aicha2012}, $\frac{R[x]}{(x^{m}-\alpha^i)}\simeq\frac{R[x]}{(x^{m}-1)}$,
$\forall i\in\{1,\ldots,2^a\}$, so then
\[
\frac{R[x]}{(x^{2^am}-1)} \simeq \prod_{i=1}^{2^a}\frac{R[x]}{(x^m-1)}.
\]
Thus, any ideal $I$ of $\frac{R[x]}{(x^{2^am}-1)}$ is equivalent to a direct sum of $2^a$ ideals $I_i$ of $\frac{R[x]}{(x^{m}-1)}$.
Therefore, a cyclic code over $R$ is a direct sum of $2^a$ cyclic codes of length $m$ over $R$.
\qed

\section{The Existence of Cyclic Isodual Codes over Finite Chain Rings}
In this section, conditions are given on the existence of cyclic isodual codes over finite chain rings.
Explicit constructions of monomial isodual cyclic free codes for odd characteristics are also provided.
We begin with the following result.
\begin{thm}
\label{prop:equivalent} Let $C$ be a cyclic code of length $n$ over
$R$ generated by the polynomial $g(x)$, and $\lambda$ a unit in $R$ such that $\lambda^n=1$.
Then the following holds.
\begin{enumerate}
\item[(i)] If $C$ is free, then $C$ is equivalent to the cyclic code generated by $g^*(x)$.
\item[(ii)] $C$ is equivalent to the cyclic code generated by $g(\lambda x)$.
\end{enumerate}
\end{thm}
\pf
Let $a=-1$ so that $(-1,n)=1$.
Then the multiplier
\begin{equation}
\begin{array}{ccl}
\label{eq:ling2}\mu_{-1}:R[x]/\langle x^n-1 \rangle &\longrightarrow
&
 R[x]/\langle x^n-1 \rangle\\
 f(x)&\mapsto & \mu_{-1}(f(x)) =
f(x^{-1}),
\end{array}
\end{equation}
is a ring automorphism.
Furthermore, $\mu_{-1}$ is a weight
preserving linear transformation for codes over finite chain rings.
If $c(x)=c_0+c_1x+c_2x^2+\ldots +c_kx^k \in C$, then $\mu_{-1}(c(x))= c(x^{-1})= x^{n-k}
(c_k+c_{k-1}x+c_{k-2}x^2+\ldots +c_0x^k)$.
This shows that the multiplier $\mu_{-1}$ is weight preserving, so from
Lemma \ref{lem:greferath} $C$ and $\mu_{-1}(C)$ are monomially equivalent codes.
Let $g(x)$ and $g'(x)$ be the generator polynomials of $C$ and
$\mu_{-1}(C)$, respectively.
Since $\mu_{-1}$ is a ring automorphism, $C$ and $\mu_{-1}(C)$ have the same
dimension, so the polynomials $g(x)$ and $g'(x)$ have the same degree.

From the definition of the reciprocal polynomial of $g(x)$, $g^*(x) \in \mu_{-1}(C)$ so that $g'(x)$ divides $g^*(x)$.
For $g(0)\in R^* $, $g^*(x)$ and $g(x)$ have the same
degree so that $g^*(x)$ and $g'(x)$ also have the same degree, and
thus generate the same cyclic code.
Therefore, the free cyclic code generated by $g(x)$ is equivalent to the cyclic code generated by $g^*(x)$.

Suppose there exists $\lambda \in R^*$ such that $\lambda^n=1$ and let
\[
\begin{tabular}{cccc}
               $ \phi $: $\frac{R[x]}{(x^{n}-1)} $ &$ \longrightarrow$ & $\frac{R[x]}{(x^{n}-1)}$ \\
                     $f(x) $& $\longmapsto $& $\phi(f(x))=f(\lambda x)$. \\
                \end{tabular}
\]
For polynomials $f(x)$, $ g(x)$ $ \in R[x]$ we have that $f(x)
\equiv g(x) \bmod (x^{n}-1)$ if and only if there exists a polynomial
$h(x) \in R[x]$ such that
\[
f(x)- g(x)\,=\,h(x)\,(x^{n}-1).
\]
Thus it must be that
\[
\begin{array}{ccl}
                 f(\lambda x)-g(\lambda x) & = & h(\lambda x)[(\lambda x)^{n}-1 ] \\
                  & = &h(\lambda x)[(\lambda)^{n}x^{n}-1] \\
                  & = &h(\lambda x)[x^{n}- 1],  \\
\end{array}
\]
which is true if and only if $f(\lambda x) \equiv g(\lambda x) \bmod (x^{n}-1)$.
Thus for $f(x), g(x) \in R[x]/(x^{n}-1)$
\[
\phi(f(x)) = \phi (g(x)),
\]
if and only if
\[
g(x)= f(x),
\]
where $\phi$ is well defined and one-to-one.
It is obvious that $\phi$ is onto, and it is easy to verify that $ \phi$ is a ring
homomorphism, so $\phi$ is a ring isomorphism.
If $C=\langle g(x)\rangle$, then $\phi(C)=\langle g(\lambda x)\rangle$.
Furthermore, $\phi$ is a weight preserving linear transformation for
codes over finite chain rings.
Let $c(x)=c_0+c_1x+c_2x^2+\ldots +c_kx^k \in C$.
Since $c_i=0 \Leftrightarrow \lambda^i c_i=0$ ($\lambda^i \neq 0$),
we have that $\phi(c(x)) = c_0+\lambda c_1x+\lambda^2c_2x^2+\ldots +\lambda^kc_kx^k$.
Then the Hamming weights $\mbox{wt}(c(x))$ and $\mbox{wt}(\phi(c(x)))$ are equal, so from
Lemma \ref{lem:greferath} $C$ and $\phi(C)$ are monomially equivalent codes.
\qed

\begin{thm}
\label{thm:equivalent2} Let $R$ be a finite chain ring with residue
field $\F_q$, $q$ an odd prime power such that $q \equiv 1\bmod 2^a$,
$a$ a positive integer, $m$ an odd integer, and $f(x)$ a polynomial such that $x^m-1=(x-1)f(x)$.
Then the free cyclic codes of length $2^am$ generated by
\[
(x^{2^{a-1}}-1)\prod_{k=0}^{2^{a-1}-1} f(\alpha^{-2k-1}x),
\]
and
\[
(x^{2^{a-1}}+1)\prod_{k=1}^{2^{a-1}} f(\alpha^{-2k}x),
\]
are isodual codes of length $2^am$.
\end{thm}
\pf
By Lemma \ref{lem:primitive}, if $q\equiv 1 \bmod 2^a$,
there exists a primitive $2^a$-th root of unity $\alpha \in R^*$ such that $\alpha^{2^a}=1$.
Suppose that $x^{m}-1=(x-1)f(x)$, then
\[
(x^{2^am}-1)=(x^{2^a}-1)\prod_{k=1}^{2^a}f(\alpha ^{-k}x).
\]
Further, we have $(x^{2^a}-1)= (x^{2^{a-1}}-1)(x^{2^{a-1}}+1)$,
so that
\[
(x^{2^am}-1)=(x^{2^{a-1}}-1)(x^{2^{a-1}}+1)\prod_{k=1}^{2^a}f(\alpha^{-k}x)
\]
\[
(x^{2^am}-1)=(x^{2^{a-1}}-1)(x^{2^{a-1}}+1)\\
\prod_{k=1}^{2^{a-1}}f(\alpha ^{-2k}x)\prod_{k=0}^{2^{a-1}-1}f(\alpha^{-2k-1}x).
\]
Let
\[
g(x)=(x^{2^{a-1}}-1)\prod_{k=0}^{2^{a-1}-1} f(\alpha^{-2k-1}x),
\]
so that
\[
h(x)=(x^{2^{a-1}}+1)\prod_{k=1}^{2^{a-1}} f(\alpha^{-2k}x),
\]
and $h^*(x)=g^*(\alpha x)$.
By Theorem~\ref{prop:equivalent}(i), $C$ is equivalent to the cyclic code generated by $ g^*(x)$, and
by Theorem~\ref{prop:equivalent}(ii), the cyclic code generated by $ g^*(x)$ is equivalent
to the cyclic code generated by $g^*(\alpha x)=h^*(x)$.
As the latter code is $C^\perp$, $C$ is isodual, so then the cyclic code generated by $g(x)$ is isodual.
The same result is obtained for
\[
g(x)=(x^{2^{a-1}}+1)\prod_{k=1}^{2^{a-1}} f(\alpha^{-2k}x).
\]
\qed
\begin{ex}
Let $R=\Z_9$, $a=1$ and $m=5$ so that $n=10$. The polynomial $f(x)$ in Theorem \ref{thm:equivalent2}  is  $f(x)=x^4+x^3+x^2+x+1$.
The polynomials
\[
g_1(x)=x^5+7x^4+2x^3+7x^2+2x+8,
\]
and
\[
g_2(x)=x^5+2x^4+2x^3+2x^2+2x+1,
\]
generate isodual codes with minimum Hamming weight 4.
\end{ex}
\begin{thm}
\label{thm:equivalent3} Let $R$ be a finite chain ring with residue field $\F_q$, $q$ an
odd prime power such that $q \equiv 1\bmod 2^a$ with $a\geq 1$ an integer, $m$ an odd integer,
and $g_1(x)$, $g_2(x)$ polynomials in $R[x]$ such that $x^m-1=(x-1)g_1(x)g_2(x)$.
The free cyclic codes of length $2^am$ generated by
\[
(x^{2^{a-1}}-1)\prod_{k=1}^{2^{a-1}} g_i(\alpha^{-2k}x)\prod_{k=0}^{2^{a-1}-1}g_j(\alpha^{-2k-1}x),
\]
and
\[
(x^{2^{a-1}}+1)\prod_{k=1}^{2^{a-1}} g_i(\alpha^{-2k}x)\prod_{k=0}^{2^{a-1}-1}g_j(\alpha^{-2k-1}x),
\]
$i,j\in\{1,2\}, i\neq j$, are isodual codes of length $2^am$
over $R$ where $\alpha\in R^*$ is a primitive $2^a$-th root of unity.
\end{thm}
\pf
By Lemma~\ref{lem:primitive}, since $q\equiv 1 \bmod 2^a$,
there exists a primitive $2^a$-th root of unity $\alpha \in R^*$ such that $\alpha^{2^a}=1$.
Suppose that $x^m-1=(x-1)g_1(x)g_2(x)$, then
\[
(x^{2^am}-1)=(x^{2^a}-1)\prod_{k=1}^{2^a}g_{1}(\alpha ^{-k}x)g_{2}(\alpha ^{-k}x).
\]
Since $(x^{2^a}-1)= (x^{2^{a-1}}-1)(x^{2^{a-1}}+1)$, we have
\[
(x^{2^am}-1)=(x^{2^{a-1}}-1)(x^{2^{a-1}}+1)\prod_{k=1}^{2^a}g_{1}(\alpha^{-k}x)g_{2}(\alpha^{-k}x),
\]
\[
=(x^{2^{a-1}}-1)(x^{2^{a-1}}+1)
\prod_{k=1}^{2^{a-1}}g_{1}(\alpha ^{-2k}x)g_{2}(\alpha ^{-2k}x)
\prod_{k=0}^{2^{a-1}-1}g_1(\alpha^{-2k-1}x)g_2(\alpha^{-2k-1}x).
\]
Let
\[
g(x)=(x^{2^{a-1}}-1)\prod_{k=1}^{2^{a-1}}
g_i(\alpha^{-2k}x)\prod_{k=0}^{2^{a-1}-1}g_j(\alpha^{-2k-1}x), \,i\neq j,
\]
then the free cyclic code generated by $g(x)$ is isodual.
We also have
\[
h(x)=(x^{2^{a-1}}+1)\prod_{k=0}^{2^{a-1}-1}
g_i(\alpha^{-2k-1}x)\prod_{k=1}^{2^{a-1}}g_j(\alpha^{-2k}x),
\]
and $ h^*(x)= g^*(\alpha x)$ from Theorem~\ref{prop:equivalent}, so the cyclic code $\langle g(x)\rangle$ is isodual.
The same result is obtained for
\[
g(x)=(x^{2^{a-1}}+1)\prod_{k=1}^{2^{a-1}}
g_i(\alpha^{-2k}x)\prod_{k=0}^{2^{a-1}-1}g_j(\alpha^{-2k-1}x), \,i\neq j.
\]
\qed
\begin{ex}
Let $R=\Z_9$, $a=1, m=11$, so that $n=22$.
The factorization of $x^{11}-1$ over $\Z_9$ contains the polynomials
$g_1=x^5 + 3x^4 + 8x^3 + x^2 + 2x + 8$ and $g_2=x^5+7x^4+8x^3+x^2+6x+8$.
There are four possible isodual codes of length $22$.
Two of these codes are given by the generator polynomials
\[
g_1(x)=8x^{11}+5x^{10}+x^9+5x^7+4x^6+3x^5+6x^4+8x^3+6x+8,
\]
and
\[
g_2(x)=8x^{11}+3x^{10}+x^8+6x^7+6x^6+4x^5+4x^4+8x^2+5x+1.
\]
The minimum Hamming weight of these codes is $7$.
\end{ex}
\begin{rem}
If\[
g(x)=(x^{2^{a-1}}-1)\prod_{k=1}^{2^{a-1}}
g_i(\alpha^{-2k}x)\prod_{k=1}^{2^{a-1}}g_j(\alpha^{-2k}x), \,i\neq j,
\]
then
$g(x)=(x^{2^{a-1}m}-1)=(x^{\frac{n}{2}}-1)$,
and the free cyclic code generated by $g(x)$ is isodual.
\end{rem}

\section{Isodual Cyclic Codes over Finite Chain Rings from Duadic Codes}
The previous section gave conditions on the existence of isodual cyclic codes over finite chain rings
and constructions for these codes.
However, a more straightforward means of finding these codes is desirable.
Note that isodual codes cannot be duadic since their length is even.
We next recall some results regarding duadic codes which will be used in this section.

Let $S_1$ and $S_2$ be unions of cyclotomic cosets modulo $m$ such that
$S_1 \cap S_2 = \emptyset$, $S_1 \cup S_2 = \Z_m \setminus \{0\}$,
and $\mu_aS_i \bmod n = S_{(i+1) \bmod 2}$.
Then the triple $\mu _a, S_1, S_2$ is called a splitting modulo $m$.
The odd-like duadic codes $D_1$ and $D_2$ are the cyclic codes over $\mathbb{F}_q$ with
defining sets $S_1$ and $S_2$ and generator polynomials
$f_{1}(x)=\Pi_{i\in S_{1}}(x-\alpha^{i}) $ and $f_{2}(x)=\Pi_{i\in S_{2}}(x-\alpha^{i})$, respectively.
The even-like duadic codes
$C_1$ and $C_2$ are the cyclic codes over $\mathbb{F}_q$ with
defining sets $\{0\}\cup S_1$ and $\{0\} \cup S_2$, respectively.

\subsection{Lifts of Duadic Codes over Finite Chain Rings}
In this section $R$ is a finite chain ring with maximal ideal
$\langle \gamma \rangle$, nilpotency index $e$,
and residue field $\F_{q}$, $ q= p^{t}$.
\begin{lem}
Let $n$ be an odd integer such that $(p,n)=1$ and  $q \equiv \Box \bmod n$.
Then there exists a pair of monic factors of $x^n-1$ in $R[x]$,
$g_i(x)$, $i\in\{1,2\}$,
such that
\[
x^{n}-1 =(x-1) g_{1}(x)g_{2}(x).
\]
\end{lem}
\pf Let $n$ be an odd integer such that $(p,n)=1$ and $q \equiv\Box \bmod n$.
Then there exists a pair of odd-like duadic codes over
$\F_{q}$ generated by $f_{1}(x)$ and $f_{2}(x)$, respectively,
with $ x^{n}-1= (x-1) f_{1}(x)f_{2}(x)$ over $\F_{q}$.
Since $x-1$, $f_{1}(x)$ and $f_{2}(x)$ are monic coprime factors of
$x^{n}-1 $ over $\F_{q}$, the result follows from Lemma \ref{lem:primitive}
\qed

Let $n$ be an odd integer such that $(p,n)=1$ and $q \equiv \Box \bmod n$.
Let $g_i$, $i\in \{1,2\}$, be the lifted polynomials of
$f_i$, where the $f_i$ are generator polynomials of the duadic codes over $\F_q$.
The following definition gives the corresponding cyclic codes over $R$.

\begin{defi}
\label{def:dua}
The free cyclic codes over $R$ are
\begin{equation}
D'_{1}= \langle g_{1}(x) \rangle,
D'_{2}= \langle g_{2}(x) \rangle,
C'_{1}= \langle (x-1) g_{1}(x) \rangle,
\text{ and }
C'_{2} = \langle (x-1) g_{2}(x) \rangle,
\end{equation}
and if $e$ is even, the non free cyclic codes over $R$ are
\begin{equation}
E_{1}=\langle (x-1) g_{1}(x), \gamma^{\frac{e}{2}}g_{1}(x)g_{2}(x) \rangle,
\text{ and }
E_{2}=\langle (x-1) g_{2}(x), \gamma^{\frac{e}{2}}g_{1}(x)g_{2}(x)
\rangle.
\end{equation}
\end{defi}

In the following we give some properties of the duadic codes in Definition \ref{def:dua}.
\begin{prop}
Let $D'_i$, and $C'_i$ $i\in \{1,2\}$ be the codes given in Definition~\ref{def:dua}.
Then we have the following.
\begin{enumerate}
  \item [(i)] If the splitting is given by $\mu_{-1}$, then $D^{'\perp}_1 =C'_1$ and $D^{'\perp}_2 =C'_2$.
  \item [(ii)]If the splitting is not given by $\mu_{-1}$, then $D^{'\perp}_1 =C'_2$ and $D^{'\perp}_2 =C'_1$.
\end{enumerate}
\end{prop}
\pf
If $g(x)$ is a generator polynomial of a free cyclic code $C$ of length $n$ over $R$,
then the dual code $C^\perp$ of $C$ is the free cyclic code whose
generator polynomial is $h^*(x)$, where $h^*(x)$ is the monic reciprocal polynomial of $h(x)=(x^n-1)/g(x)$.
Then the result follows from Lemma \ref{lem:reci} and \cite[Lemma 5.1]{BGGI}.
\qed

\begin{prop}
\label{prop:matrix}
The codes $D'_{1}= \langle g_{1}(x) \rangle $ and $D'_{2}= \langle g_{2}(x) \rangle$ in Definition \ref{def:dua} are equivalent cyclic codes over $R$.
\end{prop}
\pf
Since $(n,q)=1$, $\displaystyle{\mu_a:R[x]/(x^n-1) \longrightarrow R[x]/(x^n-1)}$ defined by  $\mu_a(f(x)) = f(x^a)$
is a ring automorphism that preserves the weight.
Let $f_1(x)= \Pi_{i\in S_{1}}(x-\alpha^{i})$ where $\alpha$ is a primitive $n$-th root of unity in $\F_q$.
By Lemma \ref{lemma:reciproque}, there exists $\beta \in S$, where $S$ is a Galois extension of $R$
such that $\overline{\beta}=\alpha$ and $g_1(x)= \Pi_{i\in S_{1}}(x-\beta^{i})$.
Then $\mu_{a}(g_1(x)) = \Pi_{i\in S_{1}}(x-\beta^{ai}) = \epsilon \Pi_{j\in S_{2}}(x-\beta^{j})$
where $\epsilon$ is a unit in $R$ since the splitting is given by $\mu_a$.
Thus from Lemma \ref{lem:greferath},
$D'_{1}$ and $ D'_{2}$ are monomially equivalent cyclic codes over $R$.
\qed

\begin{lem}
Let $G$ be a generator matrix of $C'_1$ (resp. $C'_2$).
Then the following hold.
\begin{enumerate}
\item[(i)]
\begin{eqnarray}
\label{eqn:matrix} \left(
    \begin{array}{c}
      1\,1\ldots1 \\
      G \\
    \end{array}
  \right),
\end{eqnarray}
is a generator matrix of $D'_1$ (resp. $D'_2$).
\item [(ii)]
\begin{eqnarray}
\label{eqn:matrix2} \left(
    \begin{array}{c}
       G \\
      \gamma^{\frac{e}{2}}\,\gamma^{\frac{e}{2}}\ldots \gamma^{\frac{e}{2}} \\
    \end{array}
  \right),
\end{eqnarray}
is a generator matrix of $E_1$ (resp. $E_2$).
\end{enumerate}
\end{lem}
\pf For part (i), we know that $D'_1$ and $C'_1$ are cyclic codes of length
$n$ over $R$ with generator polynomials $g_{1}(x)$ and $(x-1)g_{1}(x)$, respectively.
Since $(x-1)$, $g_{1}(x)$ and
$g_{2}(x)$ are pairwise coprime over $R$, there are polynomials
$a(x)$ and $b(x)$ in $R[x]$ such that
\[
a(x)g_{2}(x)g_{1}(x)+ b(x)(x-1)g_{1}(x) = g_{1}(x).
\]
Therefore
\[
a(x)(x^{n-1}+x^{n-2}+\ldots +x+1)+b(x)(x-1)g_{1}(x)=g_{1}(x),
\]
so (\ref{eqn:matrix}) is a generator matrix of $D'_1$.
A similar result holds for $D'_2$ with $G$ a generator matrix of $C'_2$.

For part (ii), we first prove that $\langle \gamma^{\frac{e}{2}} \rangle\nsubseteq C_1'$,
where $C_1'$ is the cyclic code of length $n$ generated by $(x-1)g_1(x)$ over $R$.
The codeword $\gamma^{\frac{e}{2}}(1^n)$ can be expressed as the polynomial
$\gamma^{\frac{e}{2}} +\gamma^{\frac{e}{2}} x + \gamma^{\frac{e}{2}} x^2+\ldots + \gamma^{\frac{e}{2}} x^{n-1}$.
Substituting $x=1$ into this polynomial,
we obtain $n \gamma^{\frac{e}{2}} \neq 0 $ since the characteristic of $R$ is prime to $n$.
Therefore
$\gamma^{\frac{e}{2}} +\gamma^{\frac{e}{2}} x + \gamma^{\frac{e}{2}} x^2+\ldots + \gamma^{\frac{e}{2}} x^{n-1}$
is not a multiple of $x-1$, so that $\langle \gamma^{\frac{e}{2}} \rangle\nsubseteq C_1'$.
It follows that $E_1$ has generator matrix (\ref{eqn:matrix2}), where $G$ is a generator matrix of $C'_1$.
A similar result holds for $E_2$ with $G$ a generator matrix of $C'_2$.
\qed

\begin{rem}
Since $D_1'$ and $D_2'$ are monomially equivalent codes, from
Proposition~\ref{prop:matrix}, $E_1$ and $E_
2$ are also monomially equivalent cyclic codes.
\end{rem}

\begin{thm}
With the previous notation the following hold.
\begin{enumerate}
\item[(i)] If the splitting is given by $\mu_{-1}$, then $E_{1}$ and $E_{2}$ are self-dual.
\item[(ii)] If the splitting is left invariant by $\mu_{-1}$, then $E_{1}$ and $E_{2}$ are isodual cyclic codes over $R$.
\end{enumerate}
\end{thm}
\pf Let $f_i, i\in\{1,2\}$, be the generator polynomials of the
odd-like duadic codes over $\F_{q}$ of length $n$. Then we have
$x^{n}-1 =(x-1)f_{1}(x)f_{2}(x)$ over $\F_{q}$. If the splitting is
given by $\mu_{-1}$ then $ f_{1}^*(x)=\epsilon f_{2}(x)$ and $
f_{2}^*(x)=\epsilon f_{1}(x)$. Hence by~Lemma~\ref{lem:reci}
their lifts have the same properties so that
\[
g_{1}^*(x)=\alpha g_{2}(x)\text{ and } g_{2}^*(x)= \alpha g_{1}(x),
\]
with $\alpha$ a unit in $R$ such that $\overline{\alpha}=\epsilon$.
Then for
\[
E_{1}=\langle (x-1) g_{1}(x), \gamma^{\frac{e}{2}}g_{1}(x)g_{2}(x) \rangle,
\]
by Theorem~\ref{th:prince} we have that
\[
E_{1}^\perp =\langle (x-1)^* g_{2}^*(x), \gamma^{\frac{e}{2}}g_{1}^*(x)g_{2}^*(x) \rangle = \langle (x-1) g_{1}(x), \gamma^{\frac{e}{2}}g_{1}(x)g_{2}(x) \rangle,
\]
so $E_1$ is self-dual.
A similar result holds for $E_2$.

If the splitting is not given by $\mu_{-1}$, then
$f_{1}^*(x)=\epsilon f_{1}(x)$ and $ f_{2}^*(x) =\epsilon f_{2}(x)$.
Hence by~Lemma~\ref{lem:reci} their lifts have the same properties, so that
$g_{1}^*(x)= \alpha g_{1}(x)$ and $g_{2}^*(x)= \beta g_{2}(x)$,
where $\alpha$ and $\beta$ are units in $R$.
Then for
\[
E_{1}=\langle (x-1) g_{1}(x), \gamma^{\frac{e}{2}}g_{1}(x)g_{2}(x) \rangle,
\]
by Theorem~\ref{th:prince} we have that
\[
E_{1}^\perp =\langle (x-1)^* g_{2}^\star(x), \gamma^{\frac{e}{2}}g_{1}^*(x)g_{2}^*(x) \rangle = \langle (x-1) g_{2}(x),
\gamma^{\frac{e}{2}}g_{1}(x)g_{2}(x) \rangle = E_2,
\]
so $E_1$ and $E_2$ are duals of each other over $R$.
Since they are monomially equivalent, they are isodual cyclic codes over $R$.
\qed
\begin{ex}
For $n = 11\equiv -1 \bmod 4 $ and $ q= 3 \equiv \square \bmod 11$, there exists a pair of odd-like duadic codes over $\F_3$
generated by $f_1(x)$ and $f_2(x)$, respectively.
Let $g_1(x)$ and $ g_2(x)$ be the corresponding Hensel lifts over $\Z_9$.
We have the factorization
\[
x^{11}-1
=(x-1)(x^{5}+3x^{4}+8x^{3}+x^{2}+2x-1)(x^{5}-2x^{4}-x^{3}+x^{2}-3x-1),
\]
over $\Z_9$, so for $g_1(x) =x^{5}+3x^{4}+8x^{3}+x^{2}+2x-1$ we have
$g_1^*(x)=-(x^{5}-2x^{4}-x^{3}+x^{2}-3x-1) =-g_2(x)$.
Therefore
\[
C=\langle(x-1)g_i(x),3g_i(x) g_j^*(x)\rangle
\]
is a self-dual code.
\end{ex}
\begin{ex}
For $n = 31\equiv -1 \bmod 4 $ and $q= 2 \equiv \square \bmod 31$,
there exists a pair of odd-like duadic codes over $\F_2$
generated by $f_1(x)$ and $f_2(x)$, respectively.
Let $g_1(x)$ and $ g_2(x)$ be the corresponding Hensel lifts over $\Z_4$.
We have the factorization
\[
\begin{array}{ccl}
x^{31}-1
&=&(x-1)(x^5 + 3x^2 + 2x + 3)(x^5 + 2x^4 + 3x^3 + x^2 + 3x + 3)\\
(x^5 + 3x^4 + x^2 + 3x + 3)(x^5 + 2x^4 + x^3 + 3)\\
(x^5 + x^4 + 3x^3 + x + 3)(x^5 + x^4 + 3x^3 + x^2 + 2x + 3),
\end{array}
\]
over $\Z_4$, so for
$g_1(x) =(x^5 + 3x^2 + 2x + 3)(x^5 + 2x^4 + 3x^3 + x^2 + 3x + 3)(x^5 + 3x^4 + x^2 + 3x + 3)$
we have
$g_1^*(x)=-(x^5 + 2x^4 + x^3 + 3)(x^5 + x^4 + 3x^3 + x + 3)(x^5 + x^4 + 3x^3 + x^2 + 2x + 3) =-g_2(x)$.
Therefore
\[
C=\langle(x-1)g_i(x),2g_i(x) g_j^*(x)\rangle,
\]
is a self-dual code.
\end{ex}
\subsection{Construction of Free Isodual Cyclic Codes over Finite Chain Rings using Lifts of Duadic Codes}

Let $n$ be an integer such that $(n,q)=1$ so $R[x]/(x^{n}-1)$ is a principal ideal ring.
The free cyclic codes over $R$ are generated by factors of $x^n-1$~\cite{G-G},
so from Theorems~\ref{prop:equivalent} and \ref{thm:equivalent2} we obtain the following theorem.

\begin{thm}
Let $R$ be a finite chain ring with residue field $\F_{q}$, and suppose
there exists a pair of odd-like duadic codes $ D_{i}= \langle f_{i}(x) \rangle$, $i=1,2$, of length $m$.
Further, let $g_i(x)\in R[x]$ be the Hensel lift of $f_i(x)\in \F_q[x]$.
We then have the following.
\begin{enumerate}
\item [(i)] The cyclic codes $C_{ij}$ and $C_{ij}'$ over $R$ generated by
\[
(x^{2^{a-1}}-1)\prod_{k=1}^{2^{a-1}} g_i(\alpha^{-2k}x)\prod_{k=0}^{2^{a-1}-1}g_j(\alpha^{-2k-1}x),
\]
and
\[
(x^{2^{a-1}}+1)\prod_{k=1}^{2^{a-1}} g_i(\alpha^{-2k}x)\prod_{k=0}^{2^{a-1}-1}g_j(\alpha^{-2k-1}x),
\]
$i,j\in\{1,2\}, i\neq j$, where $\alpha\in R^*$ is a primitive $2^a$-th root of unity, are isodual codes of length $2^am$.
\item [(ii)] If the splitting is given by $\mu_{-1}$, then the cyclic codes $C_{i}$ and $C_{i}'$ over $R$ generated by
\[
(x^{2^{a-1}}-1)\prod_{k=1}^{2^{a}} g_i(\alpha^{-k}x),
\]
and
\[
(x^{2^{a-1}}+1)\prod_{k=1}^{2^{a}} g_i(\alpha^{-k}x),
\]
respectively, where $\alpha\in R^*$ is a primitive $2^a$-th root of unity, are isodual codes of length $2^am$.
\item [(iii)] If the splitting is not given by $\mu_{-1}$, then
the dual of the cyclic code generated by
\[
(x^{2^{a-1}}-1)\prod_{k=1}^{2^{a}} g_i(\alpha^{-k}x),
\]
is equivalent to the cyclic code generated by
\[
(x^{2^{a-1}}+1)\prod_{k=1}^{2^{a}} g_j(\alpha^{-k}x).
\]
\end{enumerate}
 \end{thm}
\pf
For part (i), we use Theorem \ref{thm:equivalent3}.
Let $C_{ii}=\langle g_{ii}(x)\rangle=\langle(x^{2^{a-1}}-1)\prod_{k=1}^{2^{a}} g_i(\alpha^{-k}x)\rangle$.

If the splitting is given by $\mu_{-1}$ then $f_{1}^*(x)=\epsilon f_{2}(x)$ and $ f_{2}^*(x)=\epsilon f_{1}(x)$,
and by Lemma~\ref{lem:reci}, $g_{1}^*(x)=\beta g_{2}(x)$ and $ g_{2}^*(x)=\alpha g_{1}(x)$, so
\[
C_{ii}^{\perp}= \langle h_{ii}^*(x) \rangle = \langle(x^{2^{a-1}}+1)\prod_{k=1}^{2^{a}} g_i(\alpha^{-k}x)^*\rangle
= \langle (x^{2^{a-1}}-1)\prod_{k=1}^{2^{a}} g_i(\alpha^{-k}x)= \langle \beta g_{ii}(\alpha x)\rangle,
\]
where $\alpha$ and $\beta$ are units in $R$.
Therefore, $C_{ii}\,\simeq \, F_{ii}^{\perp}$.
The proof for the codes generated by $g_{ii}(x) \,=\, (x^{2^{a-1}}+1)\prod_{k=1}^{2^{a}} g_i(\alpha^{-k}x)$ is similar.

If the splitting is not given by $\mu_{-1}$, from Lemma~\ref{lem:reci}
$g_{1}^*(x)=\beta g_{1}(x)$ and $ g_{2}^*(x)=\alpha g_{2}(x)$.
Then
\[
C_{ii}^{\perp}= \langle h_{ii}^*(x) \rangle = \langle(x^{2^{a-1}}+1)\prod_{k=1}^{2^{a}} g_i(\alpha^{-k}x)^*\rangle
= \langle (x^{2^{a-1}}-1)\prod_{k=1}^{2^{a}} g_i(\alpha^{-k}x)= \langle \beta g_{jj}(\alpha x)\rangle,
\]
where $\alpha$ and $\beta$ are units in $R$.
Therefore, $F_{ii}\,\simeq \, F_{jj}^{\perp}$.
The proof for the codes generated by $g_{ii}(x) \,=\, (x^{2^{a-1}}+1)\prod_{k=1}^{2^{a}} g_i(\alpha^{-k}x)$ is similar.
\qed
\begin{ex}
For $R=\Z_{25}$, $q = 5$ and $m=11$, $5 \equiv 16 \bmod 11$, so there exist duadic codes generated by $f_i$, $1 \leq i\leq2$.
Since $11 \equiv -1 \bmod 4$, all splittings are given by $\mu_{-1}$ and we have
\[
\begin{array}{ccl}
(x^{11}-1) &=& (x-1)(x^5+17x^4+24x^3+x^2+16x+24)(x^5+9x^4+24x^3+x^2+8x+24)\\
&=& (x-1)g_1(x)g_2(x).
\end{array}
\]
Then $\langle (x-1)g_i(x)g_j(-x)\rangle,\,\,1\leq i,j\leq 2,\, i\neq j$, is an isodual cyclic code of length $22$ with minimum Hamming distance $8$,
and $\langle(x+1)g_i(x)g_i(-x)\rangle,\,\,1\leq i\leq 2$, is an isodual cyclic code of length $22$ with minimum Hamming distance $6$.
\end{ex}
%
%

\end{document}